\documentclass[prb,twocolumn,showpacs,showkeys]{revtex4}

\usepackage{color}
\usepackage{graphicx}% Include figure files
\usepackage{dcolumn}% Align table columns on decimal point
\usepackage{bm}

\def\be{\begin{equation}}
\def\ee{\end{equation}}
\def\bea{\begin{eqnarray}}
\def\eea{\end{eqnarray}}
\def\bse{\begin{subequations}}
\def\ese{\end{subequations}}

\begin{document}

\title{Theory of laser-induced demagnetization at high temperatures}
\date{\today}
\author{A. Manchon$^{1,2}$}
\author{Q. Li$^1$}
\author{L. Xu$^1$}
\author{S. Zhang$^1$}
\affiliation{$^1$Department of Physics, University of Arizona,
Tucson, AZ 85721, USA;\\
$^2$Materials Science and Engineering, Physical Science and
Engineering Division, KAUST, Saudi Arabia}
\begin{abstract}
Laser-induced demagnetization is theoretically studied by explicitly
taking into account interactions among electrons, spins and lattice. Assuming that the demagnetization processes take place during the thermalization of the sub-systems, the temperature dynamics is given by the energy transfer between the thermalized interacting baths. These energy transfers are accounted for explicitly through electron-magnons and electron-phonons interaction, which govern the demagnetization time scale.
By properly treating the spin system in a self-consistent random
phase approximation, we derive magnetization dynamic equations for
a broad range of temperature. The dependence of demagnetization
on the temperature and pumping laser intensity is calculated in
detail. In particular, we show several salient features for
understanding magnetization dynamics near the Curie temperature.
While the critical slowdown in dynamics occurs, we find that an
external magnetic field can restore the fast dynamics. We discuss
the implication of the fast dynamics in the application of heat
assisted magnetic recording.
\end{abstract}
\maketitle
%\clearpage
\section{Introduction}
Laser-induced Demagnetization \cite{beaurepaire,gif} (LID) and
Heat Assisted Magnetization Reversal \cite{Challener} (HAMR)
constitute a promising way to manipulate the magnetization direction
by optical means. While both LID and HAMR involve laser-induced
magnetization dynamics of magnetic materials, there are several
important differences. LID is usually considered as an ultrafast
process where the hot electrons excited by the laser field transfer
their energy to the spin system, causing demagnization. The
demagnetization time scale ranges from 100 femtosecond to a few
picoseconds. For HAMR, the laser field is to heat the magnetic
material up to the Curie temperature so that the large
room-temperature magnetic anisotropy is reduced to a much smaller
value and consequently, a moderate magnetic field is able to reverse
the magnetization. The time scale for the HAMR process is about
sub-nanosecond, three orders of magnitude larger compared to
LID.

LID observations have been carried out in a number of magnetic
materials including transition metals
\cite{beaurepaire,koop,KoopmansNat,carpene}, insulators \cite{osagawara},
half-metals \cite{kise,muller,CrO2} and dilute magnetic
semiconductors \cite{wangprl}. A general consensus of the
laser-induced demagnetization process is that the high energy
non-thermal electrons generated by a laser field relax their energy
to various low excitation states of the electron, spin and lattice
\cite{RMP}. The phenomenological model for this physical picture is
referred to as {\em three-temperature model}
\cite{beaurepaire,KoopmansNat,muller} where the three interacting sub-systems
(electrons, spins, lattice) are assumed thermalized individually at
different temperatures which are equilibrated according to a set of
energy rate equations. By fitting experimental data to the model,
reasonable relaxation times of the order of several hundred
femtosecond to a few picoseconds have been determined.

Various microscopic theories \cite{zhang,koop,EY,hubner,cywinski}
have been proposed to interpret these ultrafast time scales of
electron-spin and electron-lattice relaxations. Zhang and H\"ubner
\cite{zhang} proposed that the laser field can directly excite the
spin-polarized ground states to spin-unpolarized excited states in
the presence of spin-orbit coupling, i.e., the spin-flip transition
leads to the demagnetization {\em during} the laser pulse. In this
picture, the demagnetization is instantaneous ($\approx$50-150fs).
Recent numerical simulations \cite{abinitio} show that due to a few
active "hot spots", the instantaneous demagnetization is expected
for at most a few percent of the magnetization, consistently with
experimental arguments \cite{koop1}. Koopmans et al.\cite{koop,KoopmansNat}
suggested that the excited electrons lose their spins in the
presence of spin-orbit coupling and impurities or phonons, through
an "Elliot-Yafet"-type (EY) spin-flip scattering. Recent numerical
evaluations of the EY mechanism in transition metals \cite{EY} tend
to support this point of view. Alternatively, Battiato et al. \cite{Battiato} recently modelled such ultrafast demagnetization in terms of superdiffusive currents. Finally, numerical simulations of the ultrafast demagnetization based on the phenomenological Landau-Lifshitz-Bloch equation have been achieved successfully \cite{Atxitia}.

While these demagnetization mechanisms provide reasonable estimation
for the demagnetization time scales, the theories are usually
limited to the temperature much lower than the Curie temperature and/or
make no direct connection to the highly successful phenomenological
three-temperature model \cite{beaurepaire,muller,KoopmansNat}. As it has been
recently shown experimentally \cite{kise,osagawara}, most
interesting magnetization dynamics occur near the Curie temperature.

In this paper, we propose a microscopic theory of the laser-induced
magnetization dynamics under the three-temperature framework and derive the equations that govern the
demagnetization at arbitrary temperatures.  More specifically we
predict magnetization dynamics in the critical region.

The paper is organized as follows. In Sec.~\ref{s:mech}, we propose a model for
LID processes. In Sec.~\ref{s:spin}, we describe the spin system by
the Heisenberg model which is solved by using a self-consistent
random phase approximation. In Sec.~\ref{s:model}, the central
dynamic equations for the magnetization are derived. In
Sec.~\ref{s:num}, the numerical solutions of the equations are
carried out and the connection of our results with the experimental
data of LID and HAMR is made in Sec.~\ref{s:disc}. We conclude our paper in
Sec.~\ref{s:concl}.

\section{Model of LID \label{s:mech}}
\subsection{Spin loss mechanisms}
One of the keys to understand ultrafast demagnetization is to identify the mechanisms responsible for the spin memory loss. In the case of transition metal ferromagnets for example, the spin relaxation processes lead to complex spin dynamics due to the itinerant character of the magnetization. Elliott\cite{elliott} first proposed that delocalized electrons in spin-orbit coupled bands may lose their spin under spin-independent momentum scattering events (such as electron-electron or electron-impurity interaction). This mechanism was later extended to electron-phonon scattering by Yafet and Overhauser \cite{yafet}. Consequently, the spin relaxation time $\tau_s$ is directly proportional to the momentum relaxation time $\tau_p$. Whereas the electron-electron relaxation time is on the order of a few femtoseconds\cite{quinn} (fs), the electron-impurity and electron-phonon relaxation time is on the picoseconds (ps) scale. In semiconductors, bulk and structural inversion symmetry breaking as well as electron-hole interactions lead to supplementary spin relaxation mechanisms such as D'yakonov-Perel \cite{dp} and Bir-Aronov-Pikus \cite{bap} that are beyond the scope of the present study.\par

Relaxation processes also apply to collective spin excitations such as magnons. Whereas the electron-magnon interaction conserves the angular momentum, magnon-magnon interactions  and magnon-lattice interactions in the presence of spin-orbit coupling contribute to the total spin relaxation. While the former occurs on the magnon thermalization time scale\cite{callen} (100fs), the latter is however at the second order in spin-orbit coupling and is considered to occur on the 100ps time scale. Therefore, in a laser-induced demagnetization experiment, it is most probable that all the processes mentioned above take place during the thermalization time scale of the excited electrons and excited magnons.

\subsection{Demagnetization scenario}

To establish our model, we first separate the LID processes into four
steps: (i) generation of non-thermal hot electrons by laser pumping;
(ii) relaxation of these hot electrons into thermalized electrons
characterized by an electron temperature $T_e$; (iii) energy
transfer from the thermalized hot electrons to the spin and lattice
sub-systems; (iv) heat diffusion to the environment.\par

In our model, to be given below, we will take steps (i) and (ii)
infinitely fast. In the step (i), a laser pump excites a fraction of
electrons below the Fermi sea to $\approx$1.5 eV above the Fermi
level. This excitation process is of the order of a few fs. The
photo-induced electron transition is considered spin conserving and
thus does not significantly contribute to the demagnetization
although the spin-flip electron transition could occur in the
presence of the spin-orbit coupling \cite{zhang}. \par

In step (ii), the strong Coulomb
interaction among electrons relaxes these non-thermal high-energy
electrons to form a hot electron bath which may be described by a
thermalized hot electron temperature $T_e$. During this electron thermalization process, strong electron-electron interaction-induced momentum scattering in the presence of spin-orbit coupling leads to the ultrafast transfer of the spin degree of freedom to the orbital one \cite{bigotNat}. In our model, the electron thermalization is considered instantaneous and any possible femtosecond coherent processes are disregarded \cite{bigotNatMat}. Therefore, due to ultrafast (fs) momentum scattering, the thermalized hot electrons act as a spin sink. Under this approximation, the demagnetization itself, defined as the loss of spin angular momentum, takes place during the thermalization of the electron bath in the presence of (either intrinsic or extrinsic) spin-orbit coupling.

Following the definition of the {\em three-temperature} model, we assume that the system can be described in term of three interacting baths composed of laser-induced hot electrons, spin excitations of the ground state (magnons) and lattice excitations (phonons). The applicability of this assumption is discussed in Sec. \ref{sec:mat}. Therefore, the magnetic signal essentially comes from the collective spin excitation and it is assumed that the laser-induced hot electron only contribute weakly to the magnetization. Consequently, under the assumption that the spin loss occurs during the thermalization time of the electron and spin systems, the demagnetization problem reduces to tracking the energy transfer between the spin bath and the electron and phonon bathd.

Our main objective is then to understand step (iii), where the electrons at a higher temperature transfer their energy to the spin and
lattice sub-systems. Under the electron-magnon interaction, the magnons spin is transferred to the electron system, and is eventually lost through thermalization of the electron bath. Through interactions among electrons, spins and lattice, the entire system will ultimately reach a common temperature. Finally, a heat diffusion, step (iv), will expel the heat to the environment; this last step will be considered via a simple phenomenological heat diffusion
equation.

To quantitatively determine the energy transfer among electrons,
spins and lattice in the step (iii), one not only needs to know the
explicit interaction, but also the distribution of the densities of
excitations (electrons, magnons and phonons). Within the spirit of
the three temperature model, we consider that each
sub-system (electron, spin and lattice) is thermalized, i.e., one
can define three temperatures for electrons $T_e$, spins $T_s$ and
lattice $T_l$. The justification of this important assumption has been made in the previous section and can be
qualitatively summarized: 1) For the hot electrons of the order of 1eV,
the electron-electron relaxation time is $\tau_{ee}\approx10fs$,
which is about 100 times faster than the electron-spin and
electron-phonon interactions \cite{quinn}. 2) The lattice-lattice
interaction is about one order of magnitude smaller than the
electron-electron relaxation time, $\tau_{ll}\approx100fs$
\cite{pines}. 3) Multiple spin-waves processes are known to take
place in the ferromagnetic relaxation leading to so-called Suhl
instabilities\cite{callen}. The relaxation time is of the order of
$\tau_{ss}\propto \hbar/T_c\approx100fs$ at least for high energy
magnons\cite{callen} (for long wave length magnons, the lifetime
could be significantly longer). Thus, it is reasonable to assume
that the concepts of the three temperatures are approximately valid
as long as the time scale is longer than sub-picoseconds. \par

\subsection{Model Hamiltonian}

We now propose the following Hamiltonian for LID
\begin{equation}
{\hat H}= \sum_{\mu} {\hat H}_{\mu}+{\hat H}_{es}+{\hat
H}_{el}+{\hat H}_{sl},\label{eq:H}
\end{equation}
where ${\hat H}_\mu$ ($\mu=e,s,l$) are the electron, spin and
lattice Hamiltonians, and ${\hat H}_{\mu\nu}$ ($\mu\neq \nu$) are
the interaction among sub-systems. In the remaining of the article, the hat $\hat{}$ denotes an operator. Each term is explicitly described
below.

The electron system is described by a free electron model $\hat{H}_e
= \sum_{\bf k} \epsilon_{\bf k} {\hat c}_{\bf k}^+
{\hat c}_{\bf k}$ where ${\hat c}_{\bf k}^+$ $({\hat c}_{\bf k})$
represents the electron creation (annihilation) operator. The
equilibrium distribution is simply the Fermi distribution at $T_e$.
The lattice Hamiltonian $\hat{H}_l = \sum_{\bf q \lambda } \hbar
\omega_{\bf k \lambda}^p {\hat b}_{\bf k \lambda}^+ {\hat b}_{\bf k \lambda}$ is
modeled by simple harmonic oscillators where ${\hat b}_{\bf k \lambda}^+$
(${\hat b}_{\bf k \lambda}$) is the phonon creation (annihilation) operator
and $\lambda$ is the polarization of the phonon. The phonon
distribution at $T_l$ is $n_{\bf k \lambda} = [\exp(\hbar
\omega_{\bf k \lambda}^p/k_BT_l)-1]^{-1}$. The spin Hamiltonian is
modeled by the Heisenberg exchange interaction,
\begin{equation}
\hat{H}_s = -\displaystyle\sum_{<ij>} J_{ij} {\hat{\bf S}}_i \cdot {\hat{\bf S}}_j - g\mu_B H_{ex} \sum_{i} {\hat S}_i^z, \label{eq:heisenberg}
\end{equation}
where $J_{ij}$ is the symmetric exchange integral, ${\hat{\bf S}}_i$ is
the spin operator at the site $i$, and $H_{ex}$ is the external
magnetic field applied in $z$-direction. Unlike the electron and
lattice Hamiltonians, the spin Hamiltonian is not a single particle
Hamiltonian and the distribution of the spin density is neither
a fermionic nor a bosonic distribution. To describe the equilibrium
distribution of the spin system at arbitrary temperatures, we will
model the equilibrium properties of the spin system in the next
section.

The electron-lattice interaction $\hat{H}_{el}$ is taken as a
standard form \cite{pines}
\begin{eqnarray}
{\hat H}_{el}&=&\sum_{{\bf k},{\bf q} \lambda}B_{\bf q
\lambda}({\hat c}^+_{{\bf k+q}}{\hat c}_{{\bf k}}{\hat b}_{\bf q
\lambda}+{\hat c}^+_{{\bf k-q}}{\hat c}_{{\bf k}}{\hat b}^+_{\bf q
\lambda}),\label{eq:el3}
\end{eqnarray}
where the $B_{\bf q \lambda}$ is the electron-phonon coupling
constant. For acoustic phonons, the coupling constant takes a
particularly simple form \cite{pines},
\begin{equation}
B_{\bf q \lambda} =\frac{2\epsilon_F q}{3} \sqrt{\frac{\hbar}{2M N
\omega_{\bf q \lambda}^p}}.
\end{equation}
Here $\epsilon_F$ is the electron Fermi energy and $M$ is the mass
of the ion.

The electron-spin interaction $\hat{H}_{es}$ is modeled by the
conventional exchange interaction (sd Hamiltonian):
\begin{eqnarray}
\hat{H}_{es}&=&-J_{ex}\sum_{j,{\bf k}, {\bf k'}}\hat{c}_{\bf
k}^+e^{i{\bf k}\cdot{\bf r}_j}({\hat{\bm \sigma}}\cdot{\hat{\bf
S}}_j)\hat{c}_{\bf k'}e^{-i{\bf k'}\cdot{\bf r}_j},\label{eq:Hes}
\end{eqnarray}
where we have assumed a constant coupling constant $J_{ex}$ and
${\hat{\bm \sigma}}$ is the electron spin. When one replaces ${\hat{\bm
\sigma}}\cdot{\hat{\bf S}}_j$ by
${\hat\sigma}_z{\hat S}_j^z+\frac{1}{2}({\hat\sigma}_-{\hat S}_j^++{\hat\sigma}_+{\hat S}_j^-)$, the above
$H_{es}$ contains two effects: the first term is responsible for the
spin-splitting of the conduction bands and the second term leads to
a transfer of angular momentum between the spins of the hot electrons and the spins of the ground state, i.e. spin-waves generation and annihilation. While the interaction conserves the total spin angular momentum, the thermalization process of each bath is not spin conserving as mentioned above. Therefore, this interaction transfers energy between the electron and spin baths, which results in the effective demagnetization of the magnon bath. Consequently, the generation of magnons by hot electron is a key mechanism in our model (see also Ref. \onlinecite{carpene}).\par

Finally, the spin-lattice interaction $\hat{H}_{sl}$ has been
attributed to spin-orbit coupling \cite{vanvleck}. The energy
and the angular momentum conservations require $\hat{H}_{sl}$
containing two-magnon (${\hat a}_{\bf q}^+ {\hat a}_{\bf q'}$) and two-phonon
operators (${\hat b}_{\bf k}^+ {\hat b}_{\bf k'}$). Since the spin-orbit coupling
is already treated as a perturbation, this process is second
order in the spin-orbit coupling parameter and it is expected to be
rather small \cite{vanvleck}. Thus, $\hat{H}_{sl}$ is much smaller
than $\hat{H}_{es}$ and $\hat{H}_{el}$, and we place
$\hat{H}_{sl}=0$ throughout the rest of the paper.

To summarize our model, we consider three subsystems (electrons,
spins, and lattice) described by $\hat{H}_e$, $\hat{H}_s$ and
$\hat{H}_l$ respectively. These subsystems have their individual
equilibrium temperatures $T_e$, $T_s$ and $T_l$. The heat or energy
transfer among them are given by the interaction $\hat{H}_{es}$ and
$\hat{H}_{el}$. To determine the kinetic equation for three
subsystems, we should first establish the low excitation properties
of the spin system from $\hat{H}_s$ and relate $T_s$ to the
magnetization $m(T_s)$.

\subsection{Materials considerations\label{sec:mat}}

As stated in the introduction, laser-induced demagnetization has been observed in a wide variety of materials presenting very diverse band structures and magnetism. From the materials viewpoint, the present model makes three important assumptions: (i) laser-induced hot electrons, ground state spin excitations and phonons can be treated as separate interacting sub-systems; (ii) there exists a direct interaction between hot electrons and collective spin excitations; (iii) the excited spin sub-system can be described in terms of spin-waves.\par

Whereas the consideration of a separate phonon bath is common, the separation between the electron and spin populations may seem questionable. In systems where the itinerant and localized electrons can be identified (such as 4$f$-rare earth or carrier-mediated dilute magnetic semiconductors), it seems quite reasonable. However, in typical itinerant ferromagnets such as transition metals, the magnetism arises from a significant portion of itinerant electrons. We stress out that in our model, the separation between electron and spin baths arises from the fact the electrons we consider are laser-induced hot electrons near Fermi level (in the range $[\epsilon_F-k_BT_e,\epsilon_F+k_BT_e]$), whereas the spin bath describes the magnetic behavior of electrons lying well below Fermi level. The concept of spin waves used in the present article is rather general and applies to a wide range of ferromagnetic materials. Although energy dispersion may vary from one material to another, it is unlikely to have strong influence on the main conclusions of this work.\par

The interaction between hot electrons and magnons is actually more restrictive since it assumes overlap between electrons near and far below Fermi level. For example, this approach does not apply to half-metals (electron-magnon interaction is quenched by the 100\% spin polarization) or magnetic insulators. Nevertheless, in metallic materials such as transition metals and rare-earth, this interaction does not vanish and can lead to strong spin wave generation, as demonstrated by Schmidt et al. \cite{schmidt} in Fe.

\section{Equilibrium properties of the spin system\label{s:spin} }

The Heisenberg model for the spin system, Eq.~(\ref{eq:heisenberg}),
has no exact solution even in equilibrium. At low temperature, the
simplest approach is based on the spin-wave approximation which
predicts Bloch's law for the magnetization $m(T) = m_0
-B(T/T_c)^{3/2}$ where $T_c$ is the Curie temperature and $B$ is a
numerical constant \cite{Ashcroft}. As the temperature approaches
the Curie temperature, Bloch's law fails. Instead, one uses a
molecular mean field to model the magnetization. The resulting
magnetization displays a critical relation near $T_c$, i.e., $m(T)
\propto (1-T/T_c)^{1/2}$. Since we are interested in modeling the
magnetization in the entire range of temperature, we describe
below a self-consistent random phase approximation which reproduces Bloch's law at low temperatures and the mean field result at high
temperatures.

We first recall some elementary relations of these spin operators
given below, \bea {{\hat S}_{i}^{+}}={{\hat S}_{i}^{x}}+\bm i {{\hat S}_{i}^{y} },\,
{\hat S}_{i}^{-}={\hat S}_{i}^{x}-\bm i {\hat S}_{i}^{y}, \label{eq:0} \\
\left[{\hat S}_{i}^{+},{\hat S}_{i}^{-}\right]=2{\hat S}_{i}^{z},\,\left[{\hat S}_{i}^{\pm},{\hat S}_{i}^{z}\right]=
\mp {\hat S}_{i}^{\pm},\label{eq:1}
\\{\hat S}_{i}^{+} {\hat S}_{i}^{-}=S(S+1)+{\hat S}_{i}^{z}-({\hat S}_{i}^{z})^2\label{eq:2},\eea
and the spin Hamiltonian, Eq.~(\ref{eq:heisenberg}) can be rewritten
as \be {\hat H}=-\sum_{ij} J_{ij}({\hat S}_{i}^{-}{\hat S}_{j}^{+}+{\hat S}_{i}^{z}{{\hat S}_{j}^{z}})
-g\mu_B H_{ex} \sum_i {\hat S}_i^z. \label{eq:3} \ee

Our self-consistent random phase approximation treats the resulting
commutator, $\left[{\hat S}_{i}^{+},{\hat S}_{i}^{-}\right]=2{\hat S}_{i}^{z} \approx
2m(T)$ as a c-number, where $m(T)$ is the thermal average of ${\hat S}_i^z$
to be determined self-consistently. If we introduce the Fourier
transformation, ${\hat S}_{\bf k}^{\pm}=(1/N)\sum_{i} {\hat S}_{i}^{\pm}e^{-i\bf
{k}\cdot R_i}$, the above commutator reads as $[{\hat S}_{\bf k}^+, {\hat S}_{\bf
q}^-] = 2m(T)\delta_{\bf {k q}} $ and thus by introducing ${\hat a}_{\bf
k}^{\pm} \equiv {\hat S}_k^{\mp}/\sqrt{2m(T)}$, one has a standard boson
commutator relation $[{\hat a}_{\bf q}, {\hat a}_{\bf q'}^+]=\delta_{{\bf q}, {\bf
q}'}$. Similarly, we have $[\hat{H}_s ,{\hat a}_{\bf q}]=\hbar \omega_{\bf
q} {\hat a}_{\bf q}$, where
\begin{equation} \hbar \omega_{\bf q} = g\mu_B
H_{ex}+2m(T)\sum_{\bf q} [J_{0}-J({\bf q})]\label{eq:bspin}
\end{equation}
where $J({\bf q})=(1/N)\sum_{\bf <ij>} J_{ij} \exp[i{\bf q}\cdot
({\bf R}_i - {\bf R}_j)]$. With the above bosonic approximation, one
can self-consistently determine the magnetization $m(T)$ and other
macroscopic variables such as the spin energy and specific heat. A
particular simple case is for the spin-half $S=1/2$ where the
identity \be {\hat S}_z =S-{\hat S}^-{\hat S}^+ = 1/2-\sum_{\bf q} 2m(T) a_{\bf q}^+
a_{\bf q}\label{eq:sigmas} \ee immediately leads to the
self-consistent determination for $m(T)$
\begin{equation} m(T)=1/2-\frac{1}{N}\sum_{\bf q} \frac{2m(T)}{e^{\beta
\hbar \omega_{\bf q}(T)}-1}\label{eq:self}.
\end{equation}
At low temperature, one can approximately replace $m(T)$ by 1/2 in
the right-hand side of the equation and one immediately sees that
the above solution produces the well-known Bloch relation, i.e.,
$1/2-m(T) \propto T^{3/2}$. Near the Curie temperature, one expands
$e^{\beta \hbar \omega_{\bf q}} =1+\beta \hbar \omega_{\bf
q}+(1/2)(\beta \hbar \omega_{\bf q})^2$ and notice that $\omega_{\bf
q}$ is proportional to $m(T)$ at zero magnetic field, see
Eq.~({\ref{eq:bspin}). By placing this expansion into
Eq.~(\ref{eq:self}), the zero order term in $m(T)$ determines the
Curie temperature and the second order term gives the scaling $m^2
(T) \propto (T_c-T)$, i.e. the mean field result is recovered, $m(T)
\propto (1-T/T_c)^{1/2}$. Thus the self-consistent approach captures
both low and high temperature limiting cases. In fact, the Green's
function technique \cite{callen} has been developed to justify this approximation.

For the cases other than $S=1/2$, the relation between ${\hat S}_i^z$ and
the number of magnons is more complicated due to non-constant
$({\hat S}_i^z)^2$ and thus Eq.~(\ref{eq:2}) cannot immediately lead to a
self-consistent equation for $m(T)$. Instead, one needs to relate
$\langle({\hat S}_i^z)^2\rangle$ to $m(T)$ and the magnon density. Tyablikov
\cite{Tyablikov} introduces a decoupling method to approximate
$\langle({\hat S}_i^z)^2\rangle$ with $m(T)$ and the normalized number of magnons
\begin{equation} n_0 \equiv \frac{1}{N} \sum_{\bf q} \langle {\hat a}_{\bf q}^+
{\hat a}_{\bf q}\rangle = \frac{1}{N} \sum_{\bf q} \frac{1}{\exp (\beta
\omega_{\bf q} ) -1}. \end{equation} 
Here finds that, for arbitrary$S$, the self-consistent equation for determining $m(T)$ is
\begin{equation}
m(T) = \frac{(S-n_0)(1+n_0)^{2S+1} (1+S+n_0) n_0^{2S+1} }{
(1+n_0)^{2S+1}-n_0^{2S+1}}. \end{equation} By replacing $S=1/2$, the
above equation reduces to Eq.~(\ref{eq:self}). The magnetic energy
can be similarly obtained \cite{theorandmath} \bea E=E_0+
\frac{S-m(T)}{2n_0}\sum_q
\frac{\hbar\omega_{q}(0)+\hbar\omega_q}{\exp(\beta \omega_{\bf q} )
-1} \label{eq:7}\eea where $E_0$ is the ground state energy and
$\hbar \omega_q (0)$ is the spin wave energy at $T=0$. Once the
internal energy is obtained, the specific heat, $C_p=\partial
E/\partial T$, may be numerically calculated.

$m(T_s)$ is uniquely determined from Eq.~(14) or Eq.~(12) for
$s=1/2$, if the spin temperature is known. Thus, the laser-induced
demagnetization is solely dependent on the the time-dependent spin
temperature $T_s$. Before we proceed to calculate $T_s (t)$ or
$m(t)$, we show the solutions of Eq.~(14) or Eq.~(12). In Figure
\ref{fig:1}, the reduced magnetization $m(T)/S$ and the specific heat as a
function of the normalized temperature $T/T_c$ with [Figs. \ref{fig:1}(a) and (b)] and
without [Figs.~\ref{fig:1}(c) and (d)] the magnetic field are shown. A few general
features can be readily identified. First, the shapes of the
magnetization curves for different spins are very similar. Second,
the magnetic field removes the divergence of the specific heat at
the Curie temperature. As expected, the magnetization reduces to
that of the mean field result near the Curie temperature and to that
of the spin wave approximation at low temperatures.

\begin{figure}[t]
\includegraphics[width=9cm]{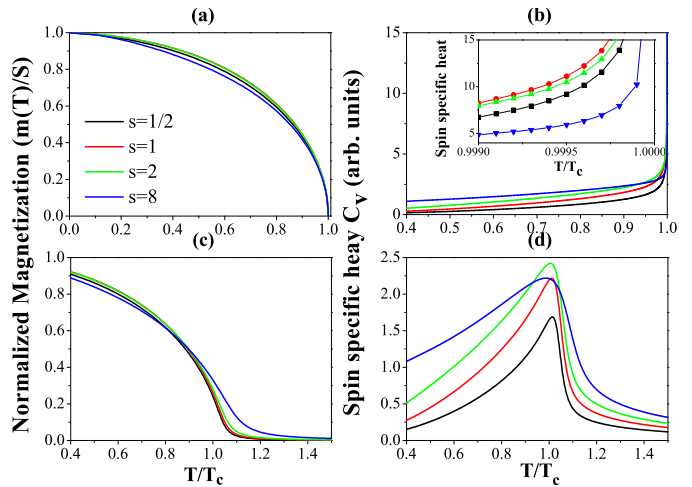}
\caption{(Color online) Temperature dependence of (a) magnetization and (b) specific heat (arbitrary unit) for
 spin=1/2,1,2,8 in the absence of the external field; Temperature dependence of (c) magnetization and (d) specific heat for spin=1/2,1,2,8 in an external field $H/T_c=0.001$.}\label{fig:1}
\end{figure}

\section{Dynamic Equations \label{s:model}}

The energy or heat transfer among electrons, spins and lattice may
be captured by the general rate equations given below,
\begin{eqnarray}\label{eq:cp1}
\frac{d\mathcal{E}_e}{dt}&=&-\Gamma_{es}-\Gamma_{el}\\\label{eq:cp2}
\frac{d\mathcal{E}_l}{dt}&=&\Gamma_{sl}+\Gamma_{el}\\\label{eq:cp3}
\frac{d\mathcal{E}_s}{dt}&=&\Gamma_{es}-\Gamma_{sl}
\end{eqnarray}
where ${\cal E}_i$ are the energy densities ($i=e, s, l$) and the
rate of the energy transfer $\Gamma_{ij}$ should be determined by
Eq.~(1). Since we have neglected the weaker interaction between
spins and lattice, we set $\Gamma_{sl}=0$ in the above equations. In
the following, we explicitly derive the relaxation rates of
$\Gamma_{el}$ and $\Gamma_{es}$ from Eqs.~(3) and (5).

\subsection{Electron-lattice relaxation $\Gamma_{el}$}

The energy transfer rate between electrons and phonons does not
involve the spin. The Fermi golden rule applied to Eq.~(3)
immediately leads to
\begin{eqnarray}
\Gamma_{el}&=&\frac{4\pi}{\hbar}\sum_{{\bf k},{\bf
q}}\hbar\omega_{\bf q}^{p}|B_{\bf q}|^2\delta(\epsilon_{\bf k}-\epsilon_{\bf k+q}+\hbar\omega_{\bf q}^{p})\times\\
&&(n_{\bf k+q}(1-n_{\bf k})(1+n_{\bf q}^{p})-n_{\bf k}(1-n_{\bf
k+q})n_{\bf q}^{p}),\nonumber
\end{eqnarray}
where the first (second) term represents the energy transfer from
(to) the electrons to (from) lattice by emitting (absorbing) a
phonon. Note that the electrons and phonons have different
temperatures; otherwise the detailed balance will make the net
energy transfer zero. The electron and phonon densities are given by
their respective equilibrium temperatures at $T_s$ and $T_l$, i.e.,
$n_{\bf k} =[\exp((\epsilon_{\bf k}-\epsilon_F)/k_BT_e) +1]^{-1}$ and $n_{\bf
q}^p =[\exp(\hbar \omega_{\bf q}^p/k_BT_l) -1]^{-1}$. We consider
polarization-independent acoustic phonons, i.e., $\omega_{\bf
q}^{p}=v_sq$ where $v_s$ is the phonon velocity. By replacing
$B_{\bf p}$ given in Eq.~(4) into Eq.~(19), we have
\begin{widetext}
\begin{equation}
\Gamma_{el}=\frac{4\pi}{\hbar}\left(\frac{2}{3}\epsilon_F\right)^2\frac{V}{(2\pi)^4}
\frac{m_e}{\hbar}\frac{m_e}{M}\int_0^{q_m} q^3dq\frac{e^{\frac{\hbar
v_sq}{k_BT_l}}-e^{\frac{\hbar v_sq}{k_BT_e}}}{e^{\frac{\hbar
v_sq}{k_BT_l}}-1}\int_{\epsilon_{\bf
q}}^{+\infty}\frac{e^{\frac{\epsilon-\epsilon_F}{k_BT_e}}d\epsilon}{(e^{\frac{\hbar
v_sq}{k_BT_e}}e^{\frac{\epsilon-\epsilon_F}{k_BT_e}}+1)(e^{\frac{\epsilon-\epsilon_F}{k_BT_e}}+1)},
\end{equation}
where we have defined the cut-off energy $\epsilon_{\bf q} \equiv
(q-\frac{2m}{\hbar^2}\hbar v_s)^2\frac{\hbar^2}{2m}$ which comes
from the $\delta$ function, and we have introduced the maximum
phonon wave number in the First Brillouin zone (this definition is
the same for magnons and Fermi wave vectors),
$q_m=k_F=(6\pi^2)^{1/3}/a_0$. By integrating over the electron
energy $\epsilon$, we obtain
\begin{equation}
\Gamma_{el}=\frac{4\pi}{\hbar}\left(\frac{2}{3}\epsilon_F\right)^2\frac{V}{(2\pi)^4}
\frac{m_e}{\hbar}\frac{m_e}{M}\int_0^{q_m} q^3dq(n_{\bf
q}^{p}(T_e)-n_{\bf q}^{p}(T_l))k_BT_e\left(\frac{\hbar
v_sq}{k_BT_e}-\ln\left(\frac{e^{\frac{\hbar v_s
q}{k_BT_e}}e^{\frac{\epsilon_{\bf q}-\epsilon_F}{k_BT_e}}+1}{e^{\frac{\epsilon_{\bf
q}-\epsilon_F}{k_BT_e}}+1}\right)\right),\label{eq:int0}
\end{equation}
\end{widetext}
where we have defined a Debye temperature, $\theta = \hbar v_s
q_m/k_B$. The last term of Eq.~(\ref{eq:int0}) can be approximated
by
\begin{equation}
\left(\frac{\hbar v_sq}{k_BT_e}-\ln\left(\frac{e^{\frac{\hbar v_s
q}{k_BT_e}}e^{\frac{\epsilon_{\bf q}-\epsilon_F}{k_BT_e}}+1}{e^{\frac{\epsilon_{\bf
q}-\epsilon_F}{k_BT_e}}+1}\right)\right)\approx\frac{\hbar v_s
q}{k_BT_e}\Theta(2k_F-p),
\end{equation}
for $\hbar \omega_{\bf q} \ll k_B T_e $, where $\Theta(x)$ is the
step function. Therefore, the relaxation rate becomes
\begin{equation}\label{eq:Gammael}
\Gamma_{el}=\frac{1}{\hbar}\left(\frac{2}{3}\epsilon_F\right)^2\frac{9\pi}{2V}
\frac{m_e}{M}\frac{\theta}{T_F}\left[G_4\left(\frac{T_e}{\theta}\right)-G_4\left(\frac{T_l}{\theta}\right)\right]\label{eq:Gacoustic}
\end{equation}
and $G_n(x)=x^{n+1}\int_0^{1/x}t^ndt/(e^t-1)$.\par

Interestingly, for $T_e,T_l\gtrsim\theta$, the relaxation rate
reduces to:
\begin{eqnarray}\label{eq:Gammaelapp}
\Gamma_{el}=\frac{1}{\hbar}\left(\frac{2}{3}\epsilon_F\right)^2\frac{9\pi}{8V}
\frac{m_e}{M} \left( \frac{T_e-T_l}{T_F} \right).
\label{eq:Goptical0}
%\Gamma_{el}^{P}&=&\frac{3\alpha'}{\hbar}(k_B\theta)^2\left(\frac{\theta}{T_F}\right)^{1/2}\frac{T_e-T_l}{T_F}\label{eq:Gpolar0}
\end{eqnarray}
Thus, the relaxation rate is simply proportional to the difference
between the electron and lattice temperatures($\Gamma_{el}\propto
T_e-T_l$); this is the assumption made in the earlier three
temperature model \cite{beaurepaire}.

\subsection{Electron-spin relaxation $\Gamma_{es}$}

The interaction between the electrons and spins given by Eq.~(5) may
be simplified by using the self-consistent random phase
approximation, i.e., we replace ${\hat S}_i^z$ by its thermal average
$m(T_s)$ and ${\hat S}_{\bf q}^{\pm} = \sqrt{2m(T_s)} a_{\bf q}^{\mp}$. The
electron-spin interaction can then be rewritten as
\begin{eqnarray}
\label{eq:es} {\hat
H}_{es}&=&-\frac{J_{ex}}{\sqrt{N}}\sqrt{2Sm(T_s)}\sum_{{\bf k}{\bf
q}}({\hat c}^+_{{\bf k}-{\bf q} \uparrow}{\hat c}_{{\bf k}\downarrow }{\hat a}_{\bf
q}+{\hat c}^+_{{\bf k}+{\bf q}\downarrow }{\hat c}_{{\bf k} \uparrow} {\hat a}_{\bf
q}^+),\nonumber\\\end{eqnarray} where we have dropped the $m(T)
\sigma_z$ term since it does not involve the energy transfer between
the electron and the spin.

The second order perturbation immediately leads to the electron-spin
relaxation:
\begin{eqnarray}
\Gamma_{es}&=&\frac{2\pi}{\hbar}\frac{2Sm(T_s)}{N}J_{ex}^2\sum_{{\bf
k},{\bf q}}\hbar\omega_{\bf q}\delta(\epsilon_{\bf k}-\epsilon_{{\bf
k}-{\bf q}}-\hbar\omega_{\bf q})\nonumber\\&\times&(n_{{\bf k}
\downarrow}(1-n_{{\bf k}-{\bf q} \uparrow})(1+n_{\bf q}^s)-n_{{\bf
k}-{\bf q} \uparrow}(1-n_{{\bf k}
\downarrow})n_{\bf q}^s)\nonumber\\
\end{eqnarray}
where $\omega_{\bf q}$ is the magnon frequency given by Eq.~(10),
the electron distribution is $n_{\bf k \sigma}=[\exp((\epsilon_{\bf
k}-\epsilon_F)/k_BT_e)+1]^{-1}$ and the magnon distribution is $n_{\bf
q}^s = [\exp(\hbar \omega_{\bf q})-1]^{-1}$. Note that the electron sub-system is considered unpolarized due to the strong spin relaxation occuring during thermalization. In Eq.~(26), the first
(second) term represents the electron emitting (absorbing) a magnon.
For the long wavelength, the magnon dispersion, Eq.~(10), is simply
$\hbar \omega_{\bf q} = \mu_B H_{ext} + \alpha k_B T_c q^2 a_0^2 $
where $\alpha\approx 1$. Following
the same procedure as the previous section, we find,
\begin{eqnarray}
\Gamma_{es}=\frac{4\pi}{\hbar}2SM(T_s)J_{ex}^2\frac{V}{(2\pi)^4}\frac{m^2}{\hbar^4}\nonumber\\\times
\int_{0}^{q_m}qdq(\hbar\omega_{\bf q})^2(n_{\bf q}^s(T_e)-n_{\bf
q}^s(T_s))
\end{eqnarray}
If the magnetic field is zero, the integration over $q$ can be
immediately carried out and we approximately have (discard the
numerical constant $\alpha$)
\begin{eqnarray}\label{eq:Gammaes}
\Gamma_{es}=\frac{(6\pi^2)^{10/3}J_{ex}^2m^3(T_s) }{2\hbar V} \left(\frac{T_c}{T_F}\right)^2\nonumber\\
\times\left[G_2\left(\frac{T_e}{D T_c}\right)-G_2\left(\frac{T_s}{D
T_c}\right)\right]
\end{eqnarray}
where $G_n(x)$ has been defined below Eq.~(23) and the temperature-dependent spin stiffness is $D=m(T_s) q_m^2
a^2$. An important result of this paper is that $\Gamma_{es}$ is
proportional to $m^3(T_s)$ and thus it is vanishingly small near the
Curie temperature. Furthermore, since $T_e/T_c$ and $T_s/T_c$ are
always comparable to 1, the electron-spin relaxation rate given by
Eq.~(28) is not proportional to $T_e-T_s$, which is quite different
from the previous three-temperature model \cite{beaurepaire}.

\subsection{Specific heats of subsystems}

As the right sides of the rate equations, Eqs.~(16)-(18), are
expressed in terms of three temperatures $T_e$, $T_s$ and $T_l$, we
need to relate the energy change of each system to the temperatures,
i.e., we should define the heat capacities $C_i$ for each subsystem
$d{\mathcal{E}_i} = C_i dT_i$. The specific heat depends on
the material details. To be more specific, we consider the Ni metal that has
been experimentally investigated most extensively.

\paragraph{Specific heat of the electrons}
In a free electron picture, the specific heat of an electron gas is
$C_e=\frac{1}{2}\pi n_ek_B(T_e/T_F)(1-3\pi^2/10(T_e/T_F)^2-...)$,
where $n_e$ is the electron density and $T_F$ is the Fermi
temperature \cite{kittel}. However, this approximation is usually
poor in the case of transition metals. In our model, we assume that the electron specific heat
remains proportional to the temperature $C_e=\gamma_eT_e$ where
$\gamma_e\approx 1.5\times 10^3$Jm$^{-3}$K$^{-2}$ is taken from
experiments \cite{heat}, which is smaller than the one assumed
earlier \cite{beaurepaire}.

\paragraph{Specific heat of lattice}
The phonon energy is derived from the Debye model, $E_{p}=\int
d^3{\bf q}\hbar\omega_{\bf q}^{p} n_{\bf q}^{p}(T)$. This yields
$C_l=3N_Ak_BF_D(T_D/T)$, where $N_A$ is the Avogadro number and
$F_D=\int_0^{T_D/T}x^4e^x/(e^x-1)^2dx$ is the Debye function. This
form of the lattice specific heat is consistent with Pawel et al.
\cite{heat}.

\paragraph{Specific heat of the spins}

We determine the spin specific heat from the numerical derivative of
the spin energy Eq.~(15), as explicitly calculated in Sec.~
\ref{s:spin}. In Fig.~\ref{fig:1}(b) and Fig.~\ref{fig:1}(d), we
have already shown the temperature dependence of the spin specific
heat with and without the external field.

\subsection{Summary}

We summarize below the dynamic equations that govern the
time-dependence of the three subsystem temperatures after the laser
pumping,
\begin{eqnarray}\label{eq:coup1}
C_e(T_e)\frac{dT_e}{dt}&=&-\Gamma_{el}(T_e,T_l)-\Gamma_{es}
(T_e,T_s)+P(t)\\\label{eq:coup2}
C_l(T_l)\frac{dT_l}{dt}&=&\Gamma_{el}
(T_e,T_l)-\frac{T_l-T_{rm}}{\tau_l}\\\label{eq:coup3}
C_s(T_s)\frac{dT_s}{dt}&=&\Gamma_{es} (T_e,T_s)
\end{eqnarray}
where we have used $d\mathcal{E}_\mu/dt = C_{\mu} (T_{\mu})
dT_{\mu}/dt$ and we have discarded the spin-phonon interaction. In
Eq.~(29) we have inserted $P(t)$ representing the initial laser
energy transfer to the electrons and in Eq.~(30), we have included a
phenomenological heat diffusion of phonons to environment which is
set at the room temperature $T_{rm}$, this term becomes significant
only at long time scale (subnanoseconds). The functions
$\Gamma_{ij}$ are:
%\begin{widetext}
\begin{eqnarray}\label{eq:Fel}
\Gamma_{el}&=&
W_{el}\left[G_4\left(\frac{T_e}{\theta_D}\right)-G_4\left(\frac{T_l}{\theta_D}\right)\right]\\\label{eq:Fes}
\Gamma_{es}&=&W_{es}m^3(T_s)\left[G_2\left(\frac{T_e}{DT_c}\right)-G_2\left(\frac{T_s}{DT_c}\right)\right]\\\label{eq:Fsl}
\end{eqnarray}
where the constants $W_{el}$ and $W_{es}$ are given in Eq.~(24) and
(28), and $D=(6\pi^2)^{1/3}m(T_s)$

\section{Numerical results\label{s:num}}

In this Section, we numerically solve our central Eqs.~(29)-(34) for
a number of plausible material parameters. Our particular focus will
be on the difference between our model and the previous
three-temperature model. Since the demagnetization is mainly
controlled by the interaction between electrons and spins, we choose
a set of different $J_{ex}$--a large $J_{ex}$ representing
transition metals (e.g., Ni, Fe and Co) and a weak $J_{ex}$ for some
ferromagnetic oxides and dilute magnetic semiconductors.  Equations
(29)-(31) are solved by using the following procedure. First, we
assume that the laser instantaneously heats the electron bath to $T_e (0)$
while the spin and lattice temperatures remain at the room
temperature $T_s(0)=T_l(0)=T_{rm}$. With these initial conditions,
we compute these temperatures after $t>0$ where the laser source
has been turned off $P(t>0)=0$. If we only consider the time
scale smaller than 100ps we may drop the heat diffusion term in
Eq.~(30).

In Fig.~\ref{fig:2}, we show the typical temperature profiles after a low
intensity laser pumping. In general, the electron-spin interaction
is stronger than the electron-phonon interaction at low temperature,
and spin and electron temperatures equilibrate within
subpicoseconds. It takes an order of magnitude longer to reach the
equilibrium between lattice and the electrons. Also shown in the
inset is the time dependent magnetization which illustrates the
fast demagnetization and slow remagnetization. In Fig.~\ref{fig:3}, we show
the temperature dependence of magnetization for different $J_{ex}$.
As expected, the demagnetization time scales with the inverse of
$J_{ex}$ while the remagnetization is independent of $J_{ex}$ since
the latter is controlled by the electron-lattice
interaction.
\par

\begin{figure}
  \includegraphics[width=9cm]{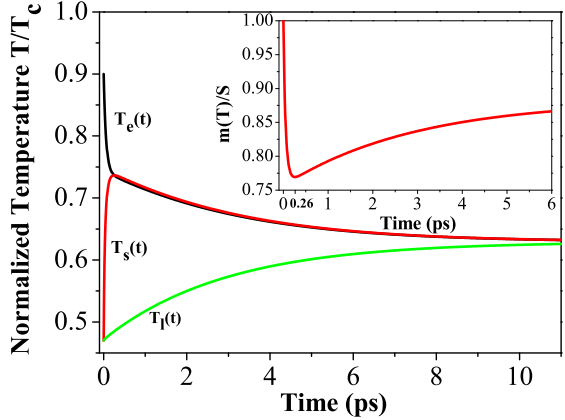}\\
  \caption{(Color online)  Time dependence of the temperatures of the electrons, spins and lattice
  after irradiation by a low intensity laser with $T_e (0)=0.7T_c$, and $T_{s}(0)=T_{p}(0)=T_{rm} =0.47T_c$,
  and $T_c=620$K. The inset shows
  the minimum magnetization (or maximum spin temperature) occurs at about 260 femtosecond.
  The other parameters are: $J_{ex}=0.15$eV, $\epsilon_F=8$eV, , $M/m=10^5$, and $a_0=0.25$nm.}\label{fig:2}

%   and large sample temperature ($T=0.85T_c$), respectively, at $P=3.5\mu$J.cm$^{-2}$;(b,d) Magnetization dynamics for
%  different pump intensities and sample temperatures, respectively. Inserts: Influence of the pump (b) and of the sample temperature
%  (c) on the demagnetization time (top insert) and amplitude (bottom insert).

\end{figure}

\begin{figure}
  \includegraphics[width=9cm]{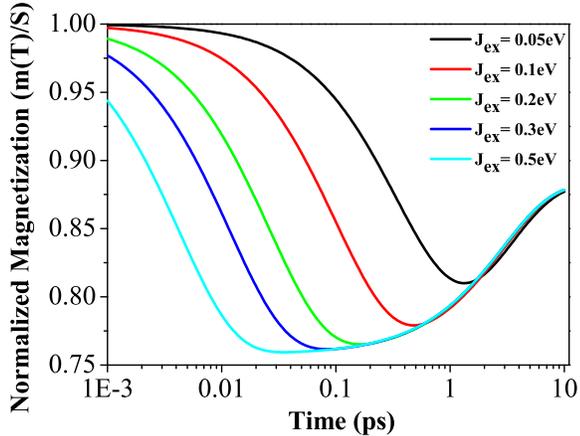}\\
  \caption{(Color online) Time-dependent magnetization as a function of time in logarithmical scale
  for various exchange parameters at a fixed laser-fluence. The parameters are same as those of Fig.~\ref{fig:2}.} \label{fig:3}
 \end{figure}
A much more interesting case is the high intensity of laser pumping.
In this case, the spin temperature raises to the Curie temperature
in 0.1-0.2 ps as shown in Fig.~\ref{fig:4}. Due to vanishingly small
magnetization at the Curie temperature, the energy transfer between
electrons and spins becomes negligible and thus the spin temperature
stays constant for an extended period of time (a
few ps). The electron temperature, however, continues to decrease
due to the electron-lattice interaction which is not affected by
the dynamic slowdown of the spins. Interestingly, after the electron
temperature drops below the spin temperature, the spin system begins
to heat the electron system and thus the electron temperature
behaves non-monotonically as seen in Fig.~\ref{fig:4}.

The dynamic slowdown of the spin temperature shown in Fig.~\ref{fig:4} is a
general property of critical phenomena. Due to the disappearance of the
order parameter (the magnetization $m(T)$ in present case), the
effective interaction reduces to zero at the critical point. In
Fig.~\ref{fig:5}, we show the time interval (labeled in Fig.~\ref{fig:5}) for the
critical slowdown as a function of the maximum spin temperature
$T_m$ for a given laser pumping power. As it is expected, the
critical slowdown shows a power law, $\tau_s \propto
[1-T_m/T_c]^{-\delta}$ with the exponent $\delta$ depending on
$J_{ex}$. In the case of very high intensity of the laser pumping,
$T_m$ can be very close to $T_c$ and the magnetization dynamics can
be extremely slow. In the presence of the external field, however,
the spin system does not have a sharp phase transition anymore and
the critical slowdown is removed, i.e., one recovers the fast
magnetization dynamics.  In Fig.~\ref{fig:6}, we compare the magnetization
dynamics with and without the magnetic field. The magnetic field
suppresses the dynamic slowdown.

\begin{figure}

  \includegraphics[width=9cm]{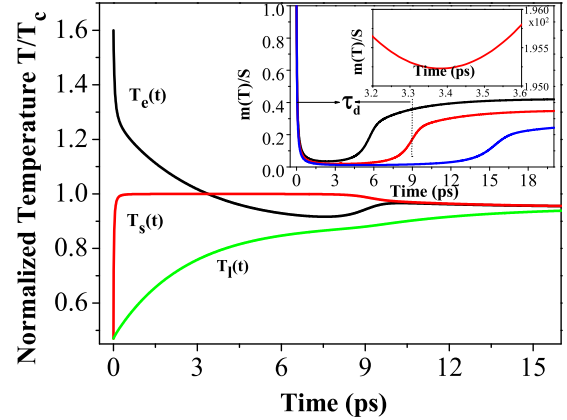}
  \caption{(Color Online) Time evolution of three temperature for a large laser-fluence case $T_e(0)=1.6$.
  The critical slowing down of the spin system is
  identified as the plateau in the figure. The inset defines a slowdown time $\tau_{d}$. The smaller inset shows
  the magnified region in the vicinity of the maximum temperature. The other parameters are same as those in Fig.~\ref{fig:2}.}\label{fig:4}
\end{figure}

\begin{figure}
  \includegraphics[width=9cm]{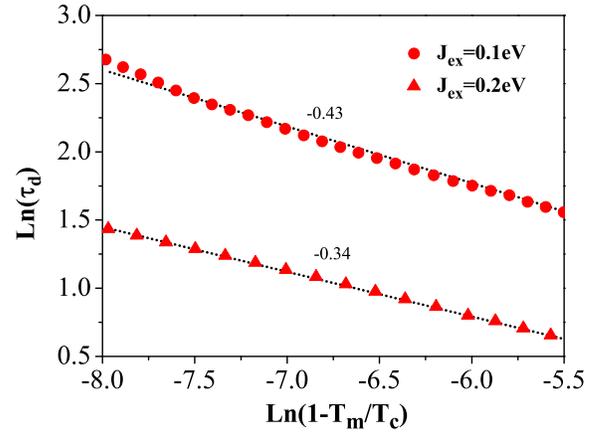}\\
  \caption{(Color online) Log-log plot of $\tau_{d}$ versus the reduced temperature for
  $J_{ex}=0.1$ and 0.2. The exponents are $\delta=0.43$ and
  $\delta=0.34$ respectively. The parameters are same as those in Fig.~\ref{fig:4}. The dashed line
  is for eye-guidance.} \label{fig:5}
 \end{figure}

\begin{figure}
  \includegraphics[width=9cm]{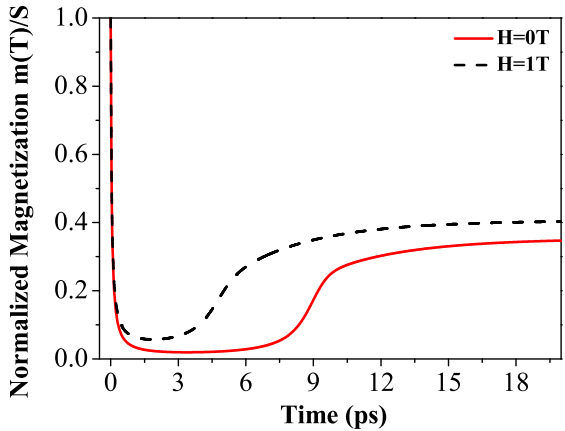}\\
  \caption{(Color online) Magnetization as a function of time with and without the magnetic
  field. The parameters are same as those in Fig.~\ref{fig:4}.
} \label{fig:6}
 \end{figure}

Finally, Fig. \ref{fig:7} shows the exponential dependence of
$\tau_{d}$ on the initial electron temperature which is directly
related to the pumping laser fluence.

\begin{figure}
  \includegraphics[width=9cm]{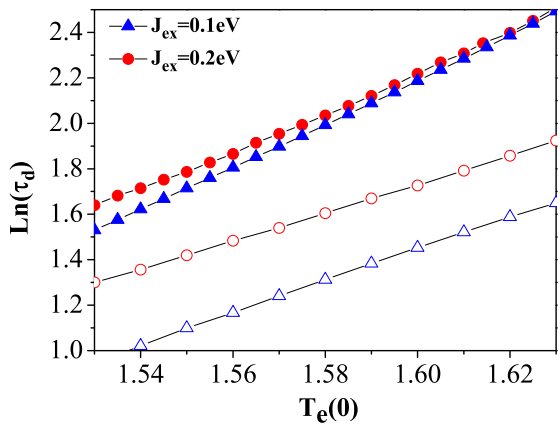}\\
  \caption{(Color online) Log($\tau_{d}$) versus the initial electron temperature for two values of $J_{ex}$ with (open symbols)
  and without (filled symbols) the magnetic field. $T_e(0)$ is normalized initial electron temperature.}
 \label{fig:7}
 \end{figure}

\section{Discussion\label{s:disc}}
\subsection{Connection with experiments on LID}

We now comment on the connection of our theory to the existing
experimental results. The LID experiments performed on transition
metallic ferromagnets \cite{beaurepaire,koop,carpene} are usually at
low laser pumping power. In these experiments, the previous
phenomenological three temperature model provides an essential
interpretation of demagnetization: the laser induced hot electrons
transfer their energies to spins and lattice. As discussed in Sec. \ref{s:mech}, in our model, the demagnetization (i.e., loss of spin memory) occurs during the instantaneous thermalization of the interacting baths. Therefore, the demagnetization/remagnetization time scale is governed by energy transfer between the baths: the demagnetization is given by the electron-spin interaction while
remagnetization time is determined by the electron-phonon interactions.\par

For transition metals, the electron-spin interaction
is at least several times larger than the electron-phonon
interaction. Thus, the demagnetization is faster than the
remagnetization. For half-metals and oxidized ferromagnets, the
demagnetization is usually longer due to a reduced electron-spin
interaction. When the temperature increases, the demagnetization
time could be significantly increased \cite{kise}; this is due to
the weakening of the effective electron-spin interaction with a reduced
magnetization. As the temperature approaches the Curie temperature,
Ogasawara et al. \cite{osagawara} observed that in all their
samples, the demagnetization time could be enhanced by one order of
magnitude.

The influence of the pump intensity on the demagnetization time can
be similarly understood. As we have shown, a large pumping intensity
creates high temperature electrons which heat the spin temperature
to the Curie temperature. Thus the temperature and the pumping
intensity dependence of the demagnetization involve the exact
physics of critical slowdown.

%Finally, we wish to briefly comment on the physical processes involved in the demagnetization and on the relation of our model to the other microscopic models. The microscopic Hamiltonian we have introduced
%conserves the total angular momenta and thus the model by itself does not lead to the demagnetization. For example, in our electron-spin interaction, the loss of the angular momentum of the spin system equals to the gain of the angular momentum of the conduction electrons. The demagnetization occurs during the {\em thermalization} of the individual systems (electrons, spins and lattice). For the electron system, the electrons relax their angular momenta to a thermal equilibrium distribution via several spin-flip processes such as the Elliott-Yafet (EY) spin flip scattering off phonons and impurities \cite{koop,KoopmansNat}. For the spin system, the thermalization is via the magnon-magnon scattering as well as the spin-orbit coupling; this latter mechanism has been emphasized by Muller et al.\cite{muller} (see also Ref. \onlinecite{bigotNat}). Since the processes due to EY and the spin-orbit coupling are typically much faster than a picosecond, they provide a strong support for our assumption of the thermalized electron and spin systems. Thus, we conclude that the energy or angular momentum transfer between the electrons and magnons (and lattice) determines the time scale of the demagnetization, though the physical process of the demagnetization comes from the other spin non-conserving interactions such as EY and spin-orbit coupling.

\subsection{Connection with HAMR}

HAMR involves heating ferromagnets to an elevated temperature so
that a moderate magnetic field is able to overcome the magnetic
anisotropy for magnetization reversal. Since the time scale in HAMR
processes is of the order of nanoseconds, the dynamics studied here
can be viewed as ultrafast, i.e., all three temperatures have
already reached equilibrium for HAMR dynamics. Even for the
temperature close to the Curie temperature, the dynamics slowdown
remains ``ultrafast'' for HAMR as long as a moderate magnetic field
is present. Thus, the HAMR dynamics could be performed in two
distinct time scales: a fast dynamics within 10 picoseconds which
determines the longitudinal magnetization $m(T)$ and a slow dynamics
from subnanoseconds to a few nanoseconds which determines the
direction of the magnetization by the conventional
Landau-Lifshitz-Gilbert equation. The detail calculations for HAMR
dynamics will be published elsewhere.

\section{Conclusion\label{s:concl}}

We have proposed a microscopic approach to the three temperature model applied to laser-induced ultrafast
demagnetization. The microscopic model consists of interactions
among laser-excited electrons, collective spin excitations and lattice. Under the assumption of instantaneous spin memory loss during the baths thermalization, the demagnetization problem reduces to energy transfer between the thermalized baths.\par

A self-consistent random phase approximation is developed to model the
low excitation of the spin system for a wide range of temperatures.
A set of dynamic equations for the time-dependent temperatures of
electrons, spins and lattice are explicitly expressed in terms of
the microscopic parameters. While the resulting equations are
similar to the phenomenological three-temperature model, there are
important distinctions in the temperature dependent properties. In
particular, the magnon softening plays a key role in demagnetization
near Curie temperature where a significant slowdown of the spin dynamics
occurs. We have also shown that for
sufficiently high temperatures (above the Debye temperature), the
dynamic properties are governed by only a few parameters: the Curie
and Fermi temperatures, the electron-spin exchange integral
$J_{ex}$, and the electron-phonon coupling constant $B_{\bf q}$. The
magnetization dynamic near the Curie temperature is rather
universal. Our numerical study of these equations illustrates that,
due to the reduction of the average magnetization as a function of
the spin temperature, both pump intensity and sample temperature are
responsible for a relative long demagnetization (several
picoseconds). An external magnetic field can suppress the critical
dynamic slowdown.

\begin{acknowledgments}
The authors acknowledge the support from DOE (DE-FG02-06ER46307) and
NSF (ECCS-1127751).
\end{acknowledgments}

\end{document}